\documentclass[aos, preprint]{imsart}

\RequirePackage[OT1]{fontenc}
\RequirePackage{amsthm,amsmath}
\RequirePackage[colorlinks,citecolor=blue,urlcolor=blue]{hyperref}
\usepackage[authoryear]{natbib}
\usepackage{graphicx}
\usepackage{makecell}
\usepackage{booktabs}
\usepackage{amsfonts}
\usepackage{graphicx}
\usepackage{enumitem}
\usepackage{multicol}
\usepackage[skip=0pt]{caption}
\usepackage{float}
\graphicspath{ {images/} }
\usepackage{algorithm}
\usepackage{algpseudocode}
\usepackage{algpascal}
\usepackage{varwidth}
\usepackage{bbold}
\usepackage{thmtools}
\usepackage{environ}
\usepackage{caption}
\usepackage{subcaption}
\usepackage{booktabs}
\usepackage{multirow}

\makeatletter
\def\BState{\State\hskip-\ALG@thistlm}
\makeatother

\arxiv{arXiv:0000.0000}

\startlocaldefs
\numberwithin{equation}{section}
\theoremstyle{plain}

\endlocaldefs

\theoremstyle{remark}

\begin{document}
	
\bibliographystyle{plainnat}
\alglanguage{pseudocode}
\begin{frontmatter}
\title{Improve Orthogonal GARCH with Hidden Markov Model }
\runtitle{Multivariate GARCH, Regime-Switching}

\begin{aug}
	\author{\fnms{Yufan} \snm{ Li,} \ead[label=e1]{} \thanksref{t1}}
		
	\thankstext{t1}{School of Engineering and Applied Sciences, Harvard University, Cambridge, USA, Email: yufan\_li@g.harvard.edu}

\end{aug}

	
		
\begin{abstract}
Orthogonal Generalized Autoregressive Conditional Heteroskedasticity model (OGARCH) is widely used in finance industry to produce volatility and correlation forecasts. We show that the classic OGARCH model, nevertheless, tends to be too slow in reflecting sudden changes in market condition due to excessive persistence of the integral univariate GARCH processes. To obtain more flexibility to accommodate abrupt market changes, e.g. financial crisis, we extend classic OGARCH model by incorporating a two-state Markov regime-switching GARCH process. This novel construction allows us to capture recurrent systemic regime shifts. Empirical results show that this generalization resolves the problem of excessive persistency effectively and greatly enhances OGARCH’s ability to adapt to sudden market breaks while preserving OGARCH's most attractive features such as dimension reduction and multi-step ahead forecasting. By constructing a global minimum variance portfolio (GMVP), we are able to demonstrate significant outperformance of the extended model over the classic OGARCH model and the commonly used Exponentially Weighted Moving Average (EWMA) model. In addition, we show that the extended model is superior to OGARCH and EWMA in terms of predictive accuracy. 
 \end{abstract}



\end{frontmatter}

\tableofcontents

\section{Introduction}
Correlations of financial returns are crucial to many financial decisions. Most portfolio optimization models require inputs of covariance matrix of sub-assets to generate optimal portfolio weights. Good correlation and volatility forecasts are therefore very important for market practitioners to make sensible investment decisions. Proposed by \cite{Ding}, \cite{AC}, Orthogonal GARCH (OGARCH) model is widely used in finance industry to generate large covariance matrix forecasts. The idea of OGARCH is to apply computation to a few key market risk factors that capture the most important uncorrelated sources of information in the original data set. In classic OGARCH, Principal Component Analysis (PCA) is applied to the data set of asset returns,  and correlation forecasts are derived by applying univariate GARCH model to the principal components. OGARCH is proved to be computationally efficient and highly accurate in forecasting the covariance matrix. However, as a passive approach it does not incorporate the possibility of sudden regime changes. As we will show later, due to the structural inflexibility, OGARCH is generally inept at reflecting dramatic market breaks such as financial crisis or abrupt policy changes. We therefore seek to develop an alternative approach based on OGARCH framework to allow detection of normal and stressed regimes so that dramatic breaks in asset correlations and volatilities can be captured in model forecasts in a more timely manner.

OGARCH model generates its forecasts based on univariate GARCH model introduced by \cite{Engle}. Univariate GARCH forecasts are based on the fact that volatility is time-varying in financial data and periods of high volatility tend to cluster. As shown in  \cite{Boll}, GARCH models are capable of providing much better volatility forecasts than constant variance models. However, forecasts of classic single-regime GARCH models are often found too high in volatile periods, as shown in \cite{Chris}. One very important reason is that GARCH tends to overstate persistent effect of large shocks in asset returns. As revealed in \cite{Gray}, the root cause for this “overstating” phenomenon is that GARCH’s structural form of conditional variances is relatively inflexible. More explicitly, GARCH estimation lends weights to past values through an autoregressive (AR) term. However, in reality, influence of past events might well vanish instantly. For example, as reported by \cite{Friedman}, large shocks to stock market returns are usually not persistent and would typically vanish rapidly. Hence, GARCH overstates volatility of the time series by artificially persisting impact of large shocks. Likewise, excessive persistence may also prevents GARCH from adapting to dramatic market breaks promptly. For example, upon arrival of financial crisis, GARCH estimation tends to reflect sudden rise in market volatility in an overly smooth and gradual fashion as it takes into account of the tranquil periods before the crash.

In classic OGARCH, dramatic breaks in asset returns (large shocks, market crashes) are also reflected in the time series of the leading principal components. In other words, structural breaks in the original data set are ``passed along'' to the principal components. While the volatility of the principal components are estimated and forecast with univariate GARCH model, OGARCH also exhibits similar ``lagging'' effect as that of the univariate GARCH models. For instance, when a sudden shock is applied to  sub-assets’ return series, classic OGARCH will first delay reflecting the shock but then leave the bump gradually fading away long after the shock has dissipated. Since OGARCH relies on the univariate GARCH to model leading principal components, it naturally inherits GARCH’s “lagging” effect in its forecasts. 

In univariate setting, it has been shown that the aforementioned issue can be effectively resolved by generalizing single-regime GARCH models to multiple regimes with different volatility levels. \cite{Hamilton} first explored the idea to use a Markov process to govern switches between different regimes. Among various Markov regime switching GARCH models proposed since, \cite{Klaassen}’s specification has the advantage of allowing multi-period-ahead forecasts and full utilization of available information. \cite{Juri} has demonstrated based on a broad set of statistical loss functions that Klaassen’s regime switching GARCH significantly outperforms usual GARCH models in forecasting volatility in short horizons. Multiple attempts have been made to introduce regime-switching to model volatility-correlation dynamics in multivariate context. Notably, Pelletier (2006) decomposes the covariances into correlations and standard deviations
and model the correlation matrix with a regime switching model, extending the classic Constant Conditional Correlation (CCC) model of Bollerslev (1990). However, unlike OGARCH, most multivariate correlation models such as Pelletier's does not allow dimension reduction. As a result, computational complexity of such models tend to grow out of hand as the data set expands.  Here, in multivariate setting, we propose a novel approach extending classic OGARCH model by combining it with Klaassen’s univariate regime-switching GARCH. By allowing an extra source of volatility persistence we will be able to enhance classic OGARCH’s flexibility to accommodate dramatic breaks while preserving OGARCH's attractive properties such as dimension reduction and multi-period ahead forecasting . We will later show that the extended regime-switching OGARCH indeed outperforms classic OGARCH, especially in the event of a sharp market regime change.

\section{Mathematical Description of the Models}
To demonstrate the superiority of the regime-switching GARCH over classic OGARCH model, we employ following competing multivariate covariance forecasting models: 
\begin{itemize}
	\item Exponentially Weighted Moving Average (EWMA) model
	\item Classic OGARCH 
	\item Markov Regime Switching OGARCH (denoted by MRSOGARCH)
\end{itemize}
We expect to see superior forecasting performance of the MRSOGARCH model comparing to others. We evaluate model forecasting performance in the context of portfolio optimization. The idea is to construct a Global Minimum Variance Portfolio (GMVP) with each model’s covariance forecasts and to see which model generates portfolio that best aligns with the specified objective, i.e. minimizing volatility. Naturally, models with better covariance forecasts will lead to better asset selection matching the pre-specified investment goal. It is also of our interest to assess predictive accuracy of the models above and to see if pairwise comparison results aligns with GMVP performance evaluation.

\subsection{EWMA model}
The Exponentially Weighted Moving Average (EWMA) model is widely used among practitioners as a simple extension to the standard weighting scheme where weights are assigned equally to past observations. EWMA assigns the largest weight to latest observation while keeping a lag term. We choose this commonly used model as an alternative to the OGARCH-based forecasting methods. 
Let $r_t$ denote vector of asset returns at day t and $\Sigma_t$ denote the covariance matrix of the underlying asset returns at day t. The covariance matrix is modeled as:
\begin{equation}
\Sigma_t=(1-\lambda) \Sigma_{t-1}+\lambda (r_{t-1}-\mu)^{T} (r_{t-1}-\mu)
\end{equation}
and 1-day ahead forecast is
\begin{equation}
\Sigma_{T+1}=(1-\lambda) \Sigma_{T}+\lambda (r_{T}-\mu)^{T} (r_{T}-\mu)
\end{equation}
where $\lambda$ is the decay factor and $T$ denotes the rolling window size for out-of-sample forecasting. i-day ahead volatility and correlation forecasts are kept as constant.

\subsection{Classic OGARCH}
In classic OGARCH model, the observed time series of $I$ underlying assets are linearly transformed to a set of uncorrelated time series (referred to as principal components) using PCA. Univariate GARCH is then applied to the principal components. 

Let $p_{it}$ denote daily closing price of asset i and $r_{it}$ denote its corresponding daily log return
\begin{equation}
r_{i,t}=log(p_{i,t})-log(p_{i,t-1})
\end{equation}
The estimation (in-sample) period consists of R observations $t=1,..,R$ and the forecasting (out-of-sample) period consists of N observation $t=R+1,..,R+N$.
Let $\bar{r}_i$ denote mean return of asset i over $t=1,..,N+R$, the normalized return of asset i at day t is calculated as
\begin{equation} 
x_{i,t}=\frac{r_{i,t}-\bar{r_i} }{v_i}
\end{equation}
where $v_i$ denotes volatility of asset i.

Let $\Sigma_p$ denote Pearson correlation matrix of normalized return of the assets. By spectral decomposition,
\begin{equation}
\Sigma_p=U \Lambda U^T
\end{equation}
where $U=[u_{ij}]\in{\mathbb{R}^{I\times I}}$ is a matrix with columns of orthonormal  eigenvectors $U_1,..,U_I$ and $\Lambda$ is the diagonal matrix of respective eigenvalues where $\lambda_1\geq,…, \geq \lambda_I$. Let $X_{t,i}=[x_{i,t}]$. The principal components $Y_i, i=1,..,I$ are obtained by
\begin{equation}
Y=XU=[Y_1,..,Y_I]
\end{equation}
The series of the i-th principal component $Y_{i,t}, i=1,..,I$ can be written as
\begin{equation}
Y_{i,t}=\epsilon_{i,t}=\eta_{i,t}\sqrt{h_{i,t}}
\end{equation}
\begin{equation}
h_{i,t}=\omega+\alpha\epsilon_{i,t-1}^2+\beta h_{i,t-1}
\end{equation}
where $\omega>0$, $\alpha \geq 0$ and $\beta \geq 0$ and innovation is expressed as the product of an $i.i.d$ process with 0 mean and unit variance times the square root of the conditional variance. 
Parameter estimation of GARCH is based on the maximum likelihood estimation on in-sample data. It can be mathematically described as
\begin{equation}
max \quad \{Loglike=\sum_{t=1}^{R} log[f(Y_{i,t})]\}
\end{equation}
where $f$ denotes the conditional probability density function of $Y_{i,t}$ per distribution assumption.
One-period ahead forecast is obtained by
\begin{equation}
h_{i,T+1}=\omega+\alpha\epsilon_{i,T}^2+\beta h_{i,T}
\end{equation}
$\tau$ period ahead forecast of volatility is found by
\begin{equation}
h_{i,T+\tau}=h+(\alpha+\beta)^{\tau} (h_{i,T}-h)
\end{equation}
where $h=\frac{\omega}{1-\alpha-\beta}$.

OGARCH aims at recovering covariance matrix of the underlying assets from variance forecasts of the the principal components. Let $U_0$ denote the matrix of the eigenvectors and $D_{T+\tau}$ denote the diagonal matrix of the variance forecasts of the corresponding principal components where $U_0 \in{\mathbb{R}^{I\times I}}$ and $D_{T+\tau} \in{\mathbb{R}^{I\times I}}$. $H_{T+\tau}$, the recovered covariance matrix ($\tau$-day ahead forecast) of normalized asset return, is obtained by
\begin{equation}
H_{T+\tau}=U_0 D_{T+\tau} U_0^T
\end{equation}
Finally, let $W$ denote the diagonal matrix where $W_{i,i}=v_i, i=1,..,I$. The covariance matrix ($\tau$-day-ahead forecast) of underlying assets, $\Sigma_{T+\tau}$, is obtained by
\begin{equation}
\Sigma_{T+\tau}=W H_{T+\tau} W
\end{equation}

\subsection{The proposed model: Markov Regime Switching OGARCH}
In classic OGARCH, leading principal components from PCA account for most variation in original data set. As GARCH is applied to time series of the leading principal components, we expect the eventual forecasts of OGARCH to inherit the “lagging” effect of the univariate GARCH model. This could significantly hinder OGARCH’s performance in the event of abrupt market regime change. Consider the case of a financial crisis. As underlying assets suffer from significant loss, we expect that the downward trend of asset returns will also be reflected in the time series of the leading components. However, due to the construct of the GARCH process (the lag term), the turbulence in the principal components’ time series is reflected in GARCH’s forecasts in a more gradual, smoother manner than it is supposed to be.  Consider assets that are tranquil in normal market condition but become highly volatile in a crisis. Fund managers relying on classic OGARCH will underestimate volatility of such assets and may not be able to react promptly. We address this limitation of OGARCH’s by employing regime-switching GARCH process in principal components' modeling, which adds one extra source of flexibility to GARCH’s forecasts. 

As stated in \cite{Juri}, the logic behind regime-switching GARCH modeling is to have a mixture of distribution characteristics and to allow the model to draw current value of the variable according to more likely states of the current observation. 

In MRSOGARCH, principal components $Y_{i,t}, i=1,..,I$,  are obtained in the same way as of classic OGARCH. However, the time series of the principal components are modeled with a Markov switching process. Depending on an unobserved process ${S_t}, t=1,..,T$, with state space $\{1,2\}$ representing turbulent and calm market conditions, each principal component $Y_{i,t}, i=1,..,I$ is characterized by 
\begin{equation}
Y_{i,t}=\mu_{i}^{(S_t)}+\epsilon_{i,t}^{(S_t)}=\mu_{i}^{(S_t)}+\eta_{i,t}\sqrt{V_{t-1}\{\epsilon_t|\tilde{S}_t\}}
\end{equation}
where $S_t=1,2$ denotes current state of the market and $\eta_{i,t}$ is a 0 mean and unit variance process. Notice that $\tilde{S}_t$ denotes regime path up to t. The state process above is modeled as a time-homogenous Markov chain with transition matrix
\[
\Pi=
\begin{bmatrix}
1-p & p \\
q & 1-q
\end{bmatrix}
\]
where $p:=\mathbb{P}(S_t=2|S_{t-1}=1)$ and $q:=\mathbb{P}(S_t=1|S_{t-1}=2)$. In Klaassen’s specification, the conditional variance is modeled as
\begin{equation}
V_{t-1}\{\epsilon_t|\tilde{S}_t\}=\omega_{S_t}+\alpha_{S_t} \epsilon_{t-1}^2+\beta_{S_t}E_{t-1} [V_{t-2}\{\epsilon_{t-1}|\tilde{S}_{t-1}\}|S_t]
\end{equation}
where $\omega_{S_t} \geq 0 $, $\alpha_{S_t} \geq 0$ and $\beta_{S_t} \geq 0$. $V_{t-1}\{\epsilon_t|\tilde{S}_t\}$ denotes the variance of $\epsilon_t$ conditional on observable information up to $t-1$ and on the regime path $\tilde{S}_t$. The expectation on the right-hand-side of the equation is across the regime path $\tilde{S}_{t-1}$ and conditional on observable information up to $t-1$ and $S_t$. It can be written out explicitly as
\begin{equation}
\begin{split}
E_{t-1} [  V_{t-2}\{\epsilon_{t-1}|\tilde{S}_{t-1}\}  |S_t=i  ]= & \tilde{p}_{ii,t-1}[(\mu^{(i)})^2+V_{t-2}\{\epsilon_{t-1}|\tilde{S}_{t-2}, S_{t-1}=i\}  ]+\\
& \tilde{p}_{ji,t-1}[(\mu^{(j)})^2+ V_{t-2}\{\epsilon_{t-1}|\tilde{S}_{t-2}, S_{t-1}=j\}  ]-\\
& [  \tilde{p}_{ii,t-1}\mu^{(i)} + \tilde{p}_{ji,t-1}\mu^{(j)}   ]^2 
\end{split}
\end{equation}
where $i,j\in{1,2}$ and $i\neq j$. The probability $\tilde{p}_{ji,t-1}$ in above expression denotes the probability of $S_{t-1}=j$ conditional on observed information up to $t-1$ and $S_t=i$, i.e. $\tilde{p}_{ji,t-1}=\mathbb{P}_{t-1}\{ S_{t-1}=j |S_t=i\}$. It can be calculated as
\begin{equation}
\tilde{p}_{ji,t-1}=\frac{p_{ji} \mathbb{P}_{t-1}(S_{t-1}=j) }{\mathbb{P}_{t-1}(S_t=i)}
\end{equation}
The ex-ante probability $ \mathbb{P}_{t-1}(S_{t}) $ can be calculated as
\begin{equation}
\mathbb{P}_{t-1}(S_{t}) =\sum_{S_{t-1=1,2}} \mathbb{P}_{t-1}(S_{t-1}) \cdot \mathbb{P}_{t-1}(S_{t}|S_{t-1})
\end{equation}
where $\mathbb{P}_{t-1}(S_{t}|S_{t-1})$ is found in the transition matrix.
The smoothed probability $ \mathbb{P}_{t-1}(S_{t-1}) $ is obtained by
\begin{equation}
\begin{split}
\mathbb{P}_{t-1}(S_{t-1}) & =\mathbb{P}_{t-2}(S_{t-1}|Y_{i,t-1})\\
& = \frac{ \mathbb{P}_{t-2}(Y_{i,t-1}|S_{t-1}) \sum_{S_{t-2=1,2}} \mathbb{P}_{t-2}(S_{t-2}) \cdot \mathbb{P}_{t-2}(S_{t-1}|S_{t-2})  }{ \mathbb{P}_{t-2}(Y_{i,t-1}) } \\
& = \frac{ \mathbb{P}_{t-2}(Y_{i,t-1}|S_{t-1}) \sum_{S_{t-2=1,2}} \mathbb{P}_{t-2}(S_{t-2}) \cdot \mathbb{P}_{t-2}(S_{t-1}|S_{t-2})  }{\sum_{S_{t-1=1,2}}  \mathbb{P}_{t-2}(Y_{i,t-1}|S_{t-1}) \mathbb{P}_{t-2}(S_{t-1}) }
\end{split}
\end{equation}
where $Y_{i,t-1}$ is the $t-1$ day observation of principal component $Y_i$.

Parameter estimation is based on maximum likelihood estimation.
\begin{equation}
max \quad \{logLike=\sum_{t=2}^{R} log[ \sum_{S_t=1,2} \mathbb{P}_{t-1}(Y_{i,t}|S_{t}) \mathbb{P}_{t-1}(S_{t}) ]\}
\end{equation}
One-day ahead forecast is found as
\begin{equation}
V_{T}\{\epsilon_{T+1}|\tilde{S}_{T+1} \}=\omega_{S_{T+1} }+\alpha_{S_{T+1} } \epsilon_{T}^2+\beta_{S_{T+1} }E_{T} [V_{T-1}\{\epsilon_{T}|\tilde{S}_{T}\}|S_{T+1} 
\end{equation}
$\tau$-day ahead forecast is found by iterating forward on
\begin{equation}
V_{t-1}\{  \epsilon_{t+i}|S_{t+i} \}=\omega_{S_{t+i}}+(\alpha_{S_{t+i}}+\beta_{S_{t+i}}) E_{t-1} [V_{t-1} \{\epsilon_{t+i-1}|S_{t+i-1}  \} |S_{t+i} ]
\end{equation}
for $i=1,..,\tau-1, \tau>1$.

Once forecasts of the volatility of the principal components are obtained, $\tau$-day-ahead covariance forecasts of original data set $\Sigma_{T+\tau}$ are recovered using equation (12) and (13). Given introduced Markov regime-switching framework, the covariance forecasts of MRSOGARCH will, as we will show later, reflect market condition change in a more timely manner and are more flexible dealing with sudden market breaks. 

\section{Discussion with Toy Examples}
In the introduction section, we have discussed the motivations behind using regime switching GARCH to model principal components of the original time series. That is, we expect the principal components to reflect broad dynamics of volatility fluctuations of the original time series, especially the volatility clustering phenomenon; and then a discrete Markov chain, representing current ``volatility state'', would provide extra freedom for adjustment from low volatility state to high volatility state. However, it is still rather vague as to how and when such improvement will take effect. In this section, we will provide some toy examples to illustrates step-by-step how MRSOGARCH works and the reasons behind. We will also highlights defects of classic GARCH models (when used along side with PCA), and what we mean exactly by ``flexibility/freedom'' in dealing with sudden volatility regime changes. 

\subsection{Bivariate Example}

\subsubsection{Model Financial Market with Two Square Waves}

This is a simplistic example but it does not entirely detach from the reality of financial markets. This example is designed to capture and exaggerate certain characteristics of financial time series so that we will see clearly how MRSOGARCH works. Consider the following square waves:

\begin{figure}[H]
	\centering
	\includegraphics[width=0.7\linewidth]{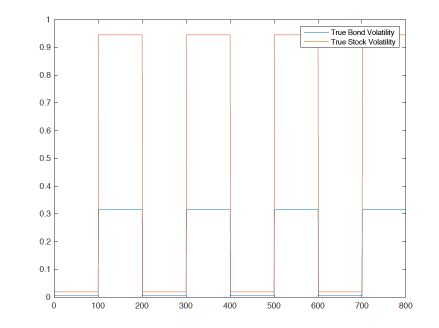}
	\caption{Square Waves Representing Two financial Assets' true Volatility}
	\label{fig:screenshot009}
\end{figure}

Let each square wave represent the true volatility of one single financial instrument. Notice that one square wave always has lower volatility than the other--the reason we do this is to mirror common portfolio construction where there are one or more less volatile assets acting as ``hedging asset', e.g. government bonds, and there are one or more assets with higher volatility such as stocks that tends to generate higher returns.  The cyclic behavior represents market dynamics of switching from normal/tranquil period (low volatility) to stressful/volatile period (high volatility). 

Now we simulate asset returns (assuming Gaussian distribution) from this volatility. 

\begin{figure}[H]
	\centering
	\includegraphics[width=0.7\linewidth]{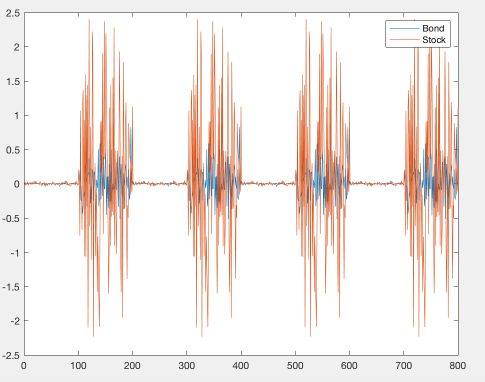}
	\caption{Simulated Asset Returns}
	\label{fig:screenshot001}
\end{figure}

The two time series are correlated with $\rho \approx 0.1$. For this example, we mostly interested in recovering the volatility series of both assets instead of their correlation structure. That is, we seek to recover Figure \ref{fig:screenshot009} as accurately as possible using both MRSOGARCH estimation and one-day forecasting.

The first step is to take the two simulated return series and decompose them into principal components. Then we shall obtain two PCA time series that are uncorrelated. We have demean and rescaled the return data o derive the principal components below (We have shown here the entire time series. Of course when actually implementing the model, this step is carried out for only in-sample data up to today). 

\begin{figure}[H]
	\centering
	\includegraphics[width=0.7\linewidth]{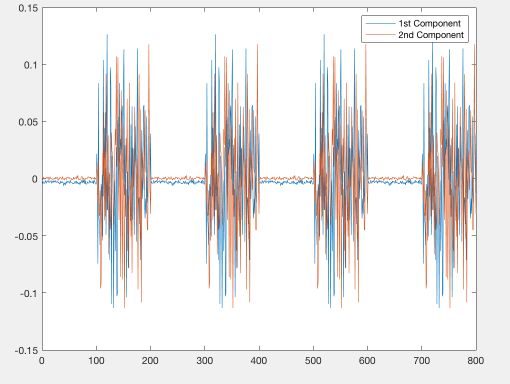}
	\caption{Two Principal Components--Rescaled}
	\label{fig:screenshot004}
\end{figure}

The main assumption based on which MRSOGARCH is motivated is that the principal components captures volatility clustering. Now we observe that it is indeed the case. The volatility of both principal components still adhere to the original cyclic pattern. The difference is that now correlation between both time series are exactly 0 so it makes sense for us to estimate and forecast each component individually. 

Now we will select our in-sample window size as 200. Note that it is important that the in-sample window include at least one volatile and one tranquil period. Otherwise the model will not be able to fully appreciate the cyclic nature of the dataset and the forecasting result will be significantly worse. We will discuss this problem further in next sections after we see the results. Now, with in-sample window size as 200, we perform maximum likelihood estimation as mentioned in last section to find suitable parameter values for each component. After finding the parameters for each component, we will keep this in-sample window size fixed and forecast one-day ahead, i.e., ``200+1-th'' day's volatility for each component. Then we move the in-sample window one step ahead and generate another forecast. And we repeat this process. 

The forecasting result for each principal component using classic GARCH(1,1) algorithm is the following:

\begin{figure}[H]
	\centering
	\includegraphics[width=0.7\linewidth]{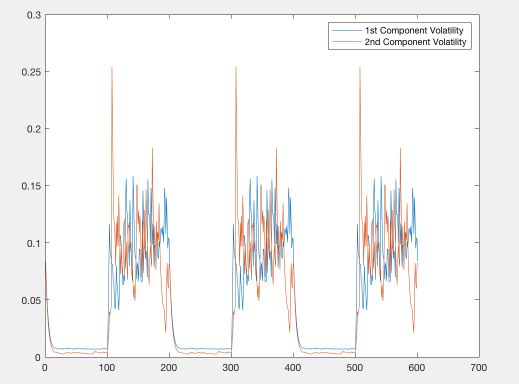}
	\caption{Forecasts on Principal Components}
	\label{fig:screenshot008}
\end{figure}

This appears to be quite messy, especially for the volatile periods. Generally, noisy fluctuation in forecasts is quite common for real-world situation. However, in our idealistic example, such noise is not at all desirable: there is simply no white noise embedded in our true volatility (two square waves as in Figure \ref{fig:screenshot009}). Indeed, we may transform this principal component forecast series back to original space/scale, we obtain the final forecast result:

\begin{figure}[H]
	\centering
	\includegraphics[width=0.7\linewidth]{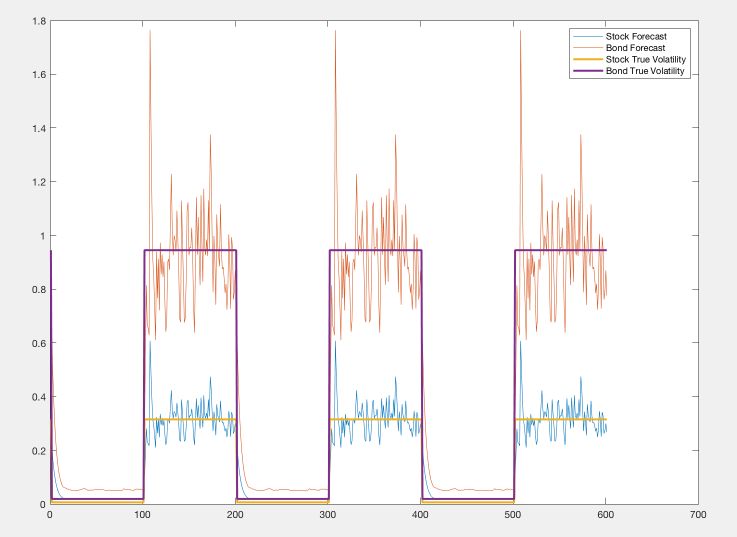}
	\caption{OGARCH Forecast Result}
	\label{fig:screenshot010}
\end{figure}

We see that the forecasts from OGARCH is not quite accurate. It is overestimating volatility for tranquil period and is quite ``undetermined'' during the volatile period. Since the final forecast result is completely determined by forecast on principal components (we are using all components in this example), the problem resides with our forecasts of principal components using GARCH(1,1). This brings us to the defects of GARCH models. To understand what is going on, recall specification of a GARCH(1,1) model:
\[\sigma_t^2=\omega+\alpha \epsilon_{t-1}^2+\beta \sigma_{t-1}^2.\]
The parameters $\omega,\alpha,\beta$ are estimated using known data (``in-sample data''). Often econometricians refer to estimation of parameters as ``training'' of the model. The significance of parameter estimation is that the model needs to learn or ``gets itself accustomed'' to characteristics of the time series it is working with. Most important to us, one of those characteristics is how much emphasis GARCH should put on latest observation $\epsilon_{t-1}$ versus all observations in the past, aggregated into $\sigma_{t-1}^2$. With volatility clustering, the more emphasis GARCH model put on latest data, the faster GARCH is at reflecting sudden changes, but at the same time it is less stable. In this example, the model is trained on the first 200 samples which include one tranquil and one volatile period: this ``forces'' GARCH model to sacrifice stability to reflect that sudden bump of volatility from tranquil period to volatile period. To see what we mean, let us decrease $\alpha$ and increase $\beta$ so that our forecast focuses more on the lag term, that is, to make our forecasts less volatile but slower:

\begin{figure}[H]
	\centering
	\includegraphics[width=0.7\linewidth]{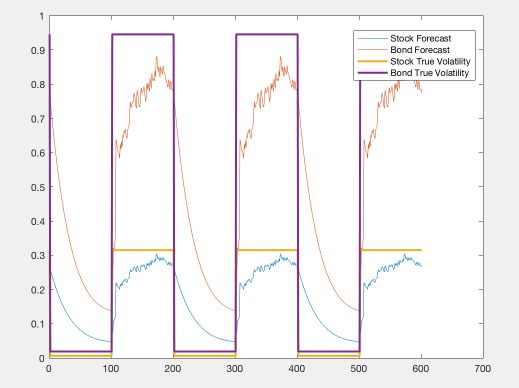}
	\caption{Slower but More stable GARCH}
	\label{fig:screenshot011}
\end{figure}

This gives us more stable/smoother estimation, but it exhibit serious ``persistency'' issue well-known for GARCH model. It is not fast enough to catch with the sudden shift of volatility regime that comes with the square wave. In fact, the overestimation of tranquil periods in \ref{fig:screenshot010} is also a result of excessive persistency; it is just less evident.

Now let's see how Markov Regime switching is able to improve upon OGARCH performance. Using MRSOGARCH, the forecasting result for each principal component is the following:

\begin{figure}[H]
	\centering
	\includegraphics[width=0.7\linewidth]{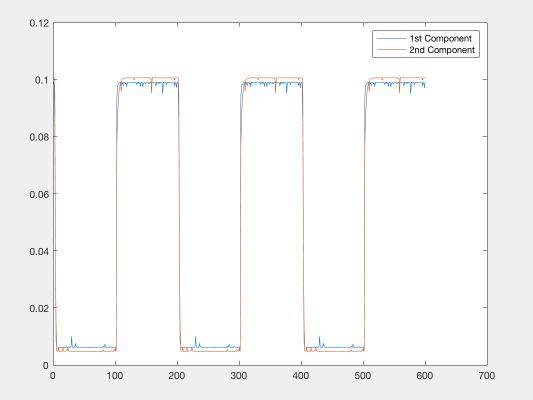}
	\caption{One-Day Ahead Forecasting Result For Principal Components}
	\label{fig:screenshot015}
\end{figure}

Comparing with, Figure \ref{fig:screenshot008}, it appears that the model is much better at handling regime switching. Indeed, revert the forecast result back to original scale gives us the final result:

\begin{figure}[H]
	\centering
	\includegraphics[width=0.7\linewidth]{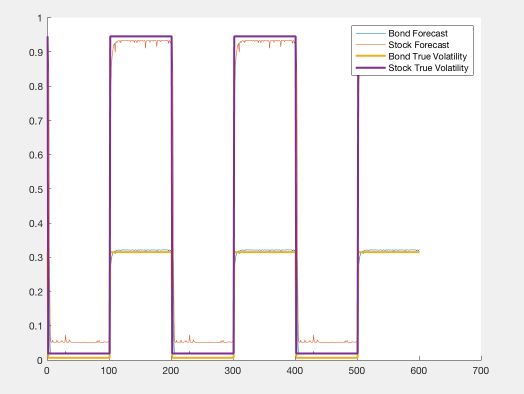}
	\caption{One Day Ahead Volatility Forecasts for Bond and Stocks}
	\label{fig:screenshot014}
\end{figure}

First, we notice that this is a much better forecast result than OGARCH. The artificial noise that presents in OGARCH forecasts is much diminished. The forecasts for volatile period is almost accurate! The only issue is that the model is still overestimating tranquil period volatility slightly and the forecasts for volatile period is a bit low. This phenomenon can be explained by looking at the ex-ante probability, i.e. the inferred probability of the current Markov state where one state represents tranquil market condition and the other state represents volatile market condition: 

\begin{figure}[H]
	\centering
	\includegraphics[width=0.7\linewidth]{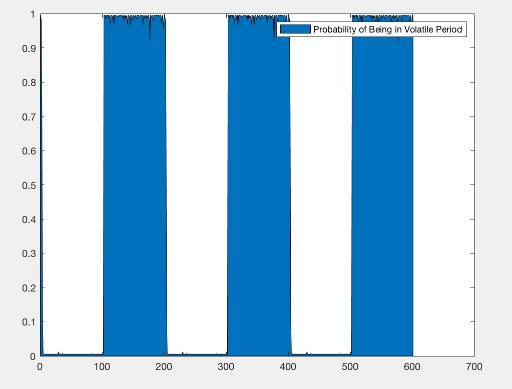}
	\caption{Probability of Being in Volatile Period}
	\label{fig:screenshot016}
\end{figure}

Notice that the inferred probability (of being in a volatile period) is quite accurate. This means that the model has been able to correctly infer which market conditions we reside within and is able to perform ``regime switching'' promptly. This is exactly how regime switching model outperforms classic GARCH: it maintains two concurrent ``streams'' of forecasts (corresponding to two states) and aggregate them weighted on inferred probability (ex-ante). Therefore, MRSOGARCH is able to switch between two regimes rapidly without sacrificing stability/accuracy. 

Though very close, inferred probability here is not exact: it was not able to be exactly 100\% sure that current state is tranquil or volatile. This is due to the fact that the model assumes that there is always a probability for us to switch to the other state, i.e. the discrete Markov chain is irreducible. And that tiny uncertainty leads to the slight underestimation and overestimation for true ``volatile volatility'' and ``tranquil volatility'' respectively. In more realistic example, this effect is negligible since there is scarcely a situation where the inferred probability should ever get that close to 1.

\subsection{Multivatiate (10-Dimension) Example and Dimension Reduction}

In 3.1, we examined the effectivity of the predictions from MSOGARCH in the bivariate setting in constrast to classic OGARCH. Now we move on to test MSORGARCH with a artificially generated 10-dimension dataset. The purpose of this example is to demonstrate that the regime switching GARCH still retains classic OGARCH's dimension reduction property, that is, several leading components are sufficient for high-dimensional application. In particular, for our artificial dataset, we found that 3 leading components are sufficient to capture most of the information meaningful to forecasts where addition of any other 7 components add little to accuracy of our forecasts. 

The dataset consists of three stages with 5000 data points in total, namely normal stage(timestamp 0-499) and volatile stage(timestamp 500-2999) and another normal stage(timestamp 3000-4999). We simulated the data in normal and volatile stages separately from different multivariate normal distributions. Both distributions have zero mean and a random sampled covariance matrix,  where elements in the covariance matrix of crisis stage are sampled from a larger range. Figure 10. is a visualization of simulated data in each dimension.

\begin{figure}[H]
	\centering
	\includegraphics[width=\linewidth]{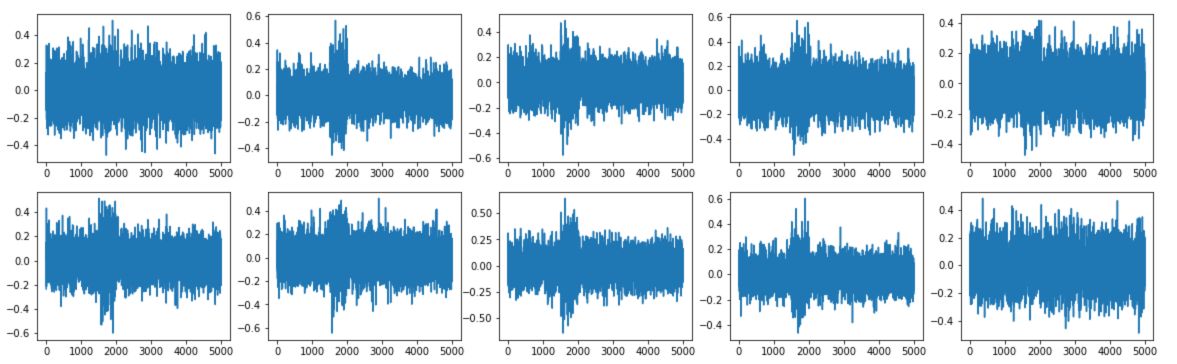}
	\caption{Visualization of 10-dimension Toy Dataset}
	\label{fig:screenshot004}
\end{figure}

We tested MSORGARCH model on 1 to 10 largest principle components. Figure 11, 12, 13, 14 shows example results with 1, 2, 5, 10 principle components. From the plots we can see that with the increasing number of principle components, the fits improve. Also, with 5 principle components, the model is able to predict both mean and fluctuation of sample variance well, compared with Figure 14 which use all 10 components. We use matrix norm,

$$
D = || \Sigma_{approx} - \Sigma_{True}|| = [\sum_{i, j}(\Sigma_{approx, i, j} - \Sigma_{True, i, j})^2]^{\frac{1}{2}}
$$

to quantify the above analysis and further understand the amount of information we lose with less components. The above defined loss increases as the difference between reconstructed covariance matrix and true covariance matrix becomes larger.  

\begin{figure}[H]
	\centering
	\includegraphics[width=\linewidth]{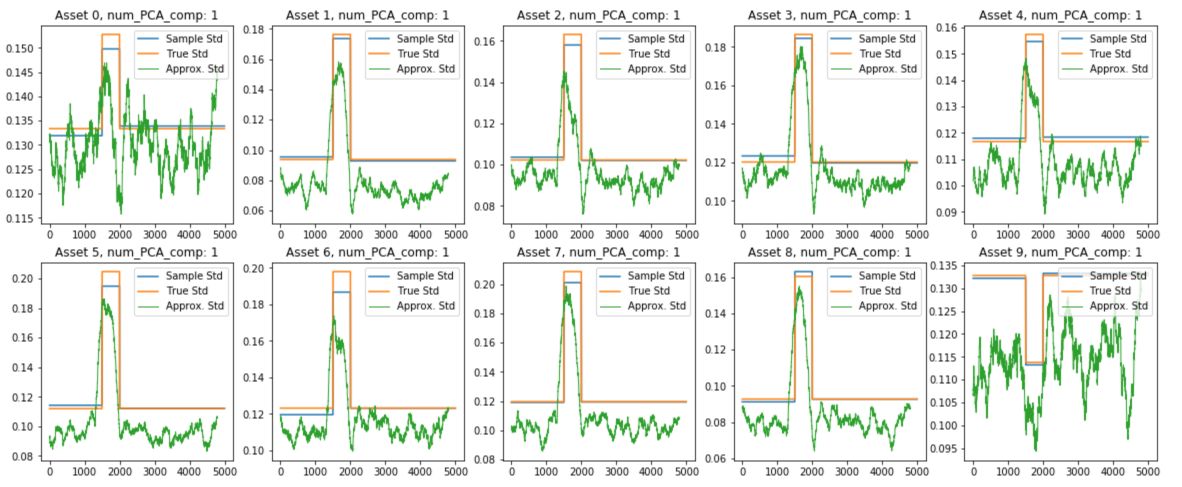}
	\caption{predict with 1 principle component}
	\label{fig:screenshot004}
\end{figure}

\begin{figure}[H]
	\centering
	\includegraphics[width=\linewidth]{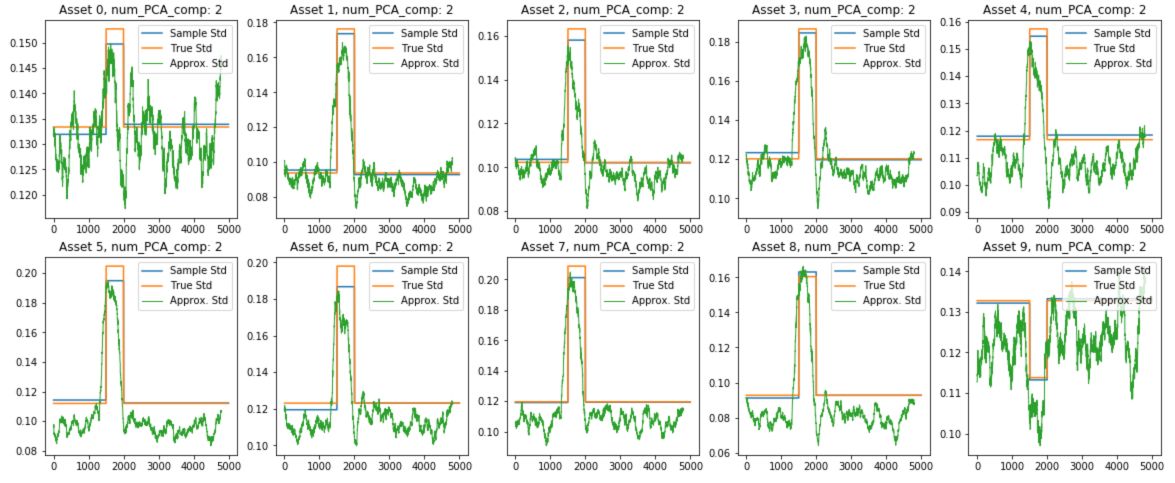}
	\caption{predict with 2 principle components}
	\label{fig:screenshot004}
\end{figure}

\begin{figure}[H]
	\centering
	\includegraphics[width=\linewidth]{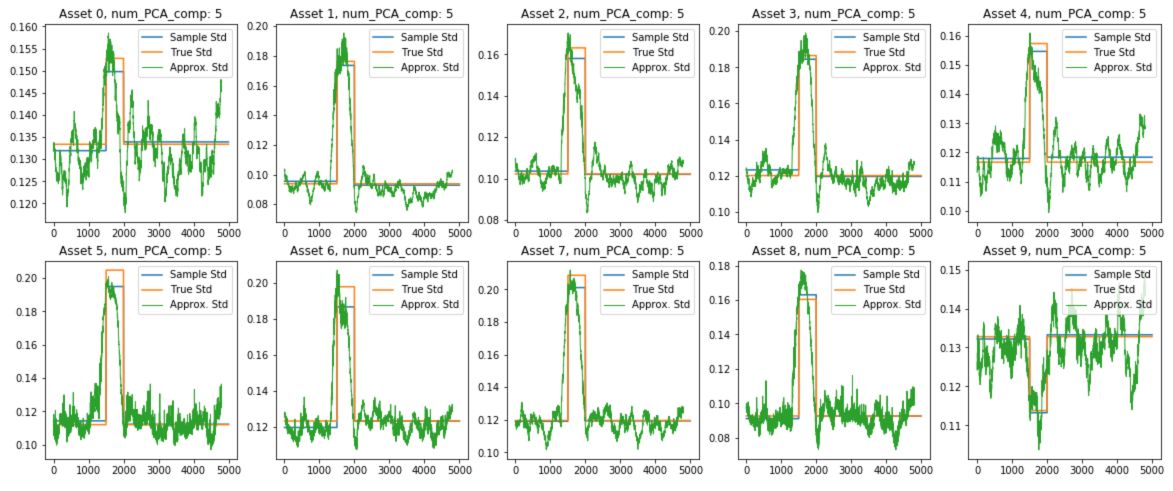}
	\caption{predict with 5 principle components}
	\label{fig:screenshot004}
\end{figure}

\begin{figure}[H]
	\centering
	\includegraphics[width=\linewidth]{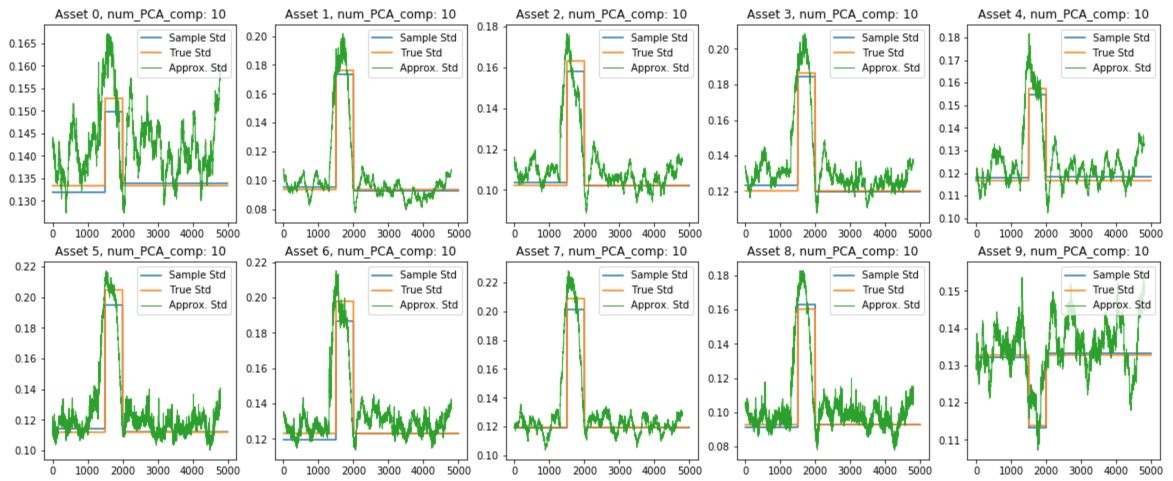}
	\caption{predict with 10 principle components}
	\label{fig:screenshot004}
\end{figure}

The matrix norm of difference between approximated covariance matrix and true covariance matrix are reported in Figure 15. We can see that under our experiment settings, there is no significant loss on D until number of required principle components are reduced to 3, which is consistent with the original OGARCH results from \cite{PCA}. Applying MSOGARCH on less principle components greatly saved computational time as well as resources without sacrificing accuracy too much. 

\begin{figure}[H]
	\centering
	\begin{tabular}{ |l|c|c|c|}
		\hline
		Number of PCA Components & $D_{total}$ & $D_{normal}$ & $D_{crisis}$ \\
		\hline
		1  & 105.13 & 77.78 & 27.35 \\
		2  & 89.12 & 65.22 & 23.90 \\
		3  & 79.83 & 57.97 & 21.85 \\
		4  & 75.35 & 54.17 & 21.17 \\
		5  & 71.00 & 50.45 & 20.54 \\
		6  & 71.20 & 50.92 & 20.28 \\
		7  & 72.40 & 51.93 & 20.47 \\
		8  & 74.20 & 53.55 & 20.65 \\
		9  & 75.91 & 55.00 & 20.91 \\
		10 & 76.97 & 55.91 & 21.06 \\
		\hline
	\end{tabular}
	\caption{D of MSORGARCH Predictions with Different Number of Principle Components}
\end{figure}

\section{A Realistic Example of Fund of Funds Management}
In this section, we will conduct a comprehensive evaluation of the merits of our regime switching model using realistic market data. We will employ different measures to evaluate the performance of the competing models. First, we compare their maximum log likelihood values. Stronger model will naturally register an increase of fit, i.e., larger MLE value. The significance of the outperformance will then be tested with the likelihood ratio test. Following \cite{Chan}, \cite{Haustch}, \cite{Amenc} etc., out-of-sample forecasting performance is evaluated by examining the economic metrics Global Minimum Variance Portfolios (GMVP)\footnote{GMVP is different from the classic Mean Variance Optimization (MVO) framework in that it avoids perils of expected return forecasting. Therefore, it is often used to assess quality of multi-asset correlation forecasting as in our case. } based on competing models' conditional covariance matrix forecasts. The idea is to assess model performance in a portfolio optimization context where the evaluation is based on quality of covariance forecasts, without being affected by noisiness and instability of mean prediction. Notice that a GMVP excludes risk-free asset and seeks to attain minimun portfolio risk. So given the same underlying assets, better covariance matrix forecasts will lead to asset allocation that lowers portfolio return volatility.

In addition, we look to evaluate model predictive accuracy. Following similar apporach as in \cite{Zangari} and \cite{Lopez}, we measure perdictive accuracy indirectly by constructing an equal-weight portfolio where daily portfolio volatility forecasts, as calculated from covariance matrix forecasts from competing models, are compared to  realized daily portfolio returns, which serve as proxies to the ``real'' daily portfolio volatility. Difference of the two time series are measured and aggregated using five loss functions while Diebold-Mariano test is applied to test significance of possible outperformance.

One key argument for using MRSOGARCH is that its constructs allow for extra source of persistence and are therefore more flexible in accommodating sudden changes in market condition. Hence, we also want to show that when such a market condition change takes place, regime switching OGARCH will be faster in recognizing it and reconstruct its portfolio most promptly. To clearly show the advantage of MRSOGARCH, we consider a portfolio of mixed asset type: 
\begin{itemize}
	\item S\&P500 Index (Bloomberg Ticker: SPX):The Standard \& Poor's 500 Index is based on the market capitalizations of 500 large companies. It is used as a proxy to equity market in this study.
	\item HFRX Equity Hedge Index (Bloomberg Ticker: HFRXEH): This index selects and rebalances components to maximize representation of the Hedge Fund Universe. It is used to represent general hedge fund performance in this study.
	\item SG Trend Index (Bloomberg ticker: NEIXCTAT): This index tracks 10 largest trend following CTAs. It is used to represent trend followers in the managed futures space in this study.
\end{itemize}
The rationale behind this selection is that during normal market condition Equity Hedge index generally has the best performance in terms of return volatility whereas SG Trend Index tend to generate most stable return during 2008 financial crisis. We expect to see that MRSOGARCH is capable of detecting this change in market regime faster, and allocate its wealth to SG Trend Index more rapidly. 

The GMVP problem can be formulated as
\[min \{\omega_{t+1:t+\tau}^T \Sigma_{t+1:t+\tau} \omega_{t+1:t+\tau}\}\]
\[s.t \quad \omega_{t+1:t+\tau}\geq 0, \omega_{t+1:t+\tau}^T 1=1\]
where $\omega_{t+1:t+\tau}$ denotes portfolio weights over investment horizon day $t+1$ to day $t+\tau$. For simplicity, we assume $\Sigma_{t+1:t+\tau}=\sum_{r=1}^{\tau} (\Sigma_{t+r})$. That is, the portfolio weight over investment horizon day $t+1$ to day $t+\tau$ is based on summation of 1-day to $\tau$-day ahead daily covariance forecasts from the model. The two constraints of the optimizer ensure a full investment of the available budget and exclude short-selling. In this study, GMVP is constructed on both 1-day and 5-day investment horizon. Notice that in a practical setting, due to the existing computing power, optimization is often run on a daily basis and portfolio weights are adjusted as the computed optimal weights drift significantly from the current asset weights. 

To evaluate the predictive accuracy of the models, we construct an equal-weight portfolio and compare volatility forecasts of the portfolio, which are based on covariance matrices from the model forecasts, and the proxy for actual portfolio volatility introduced earlier. Following \cite{Lopez}, we take absolute value of realized portfolio return over one holding period $x_{p,t+1}=|\omega_{eq}^T \cdot r_{t+1}|$ as the proxy for the actual volatility of the equal weighted portfolio. Meanwhile the variance forecasts of the equal weighted portfolio over one holding period is $h_{p,t+1}=\omega_{eq}^T \Sigma_{t+1} \omega_{eq}$. Note that $\omega_{eq}$ above denotes equal portfolio weights; $r_{t+1}$ denotes the vector of realized portfolio return over holding period $t+1$; $\Sigma_{t+1}$ denotes covariance forecast for holding period $t+1$ using information up to holding period $t$. The predictive accuracy is evaluated based on both 1-day and 5-day holding period.

We then employ five loss functions to evaluate predictive accuracy of the models. Let n denote total number of investment periods and p denote the model.

\begin{equation}
MSE_1=\frac{1}{N} \sum_{j=1}^{N} (x_{p,j}-h_{p,j}^{1/2})^2
\end{equation}
\begin{equation}
MSE_2=\frac{1}{N} \sum_{j=1}^{N} (x_{p,j}^{2}-h_{p,j})^2
\end{equation}
\begin{equation}
MAD_1=\frac{1}{N} \sum_{j=1}^{N} |(x_{p,j}-h_{p,j}^{1/2})|
\end{equation}
\begin{equation}
MAD_2=\frac{1}{N} \sum_{j=1}^{N} |(x_{p,j}^{2}-h_{p,j})|
\end{equation}
\begin{equation}
R2LOG_1=\frac{1}{N} \sum_{j=1}^{N}[log(x_{p,j}^{2} h_{p,j}^{-1})]^2
\end{equation}

where criteria (23) and (24) are classic Mean Square Errors (MSE). (25) and (26) are Mean Absolute Deviation (MAE) and are considered more robust to presence of outliers than MSE criteria as pointed out in \cite{Juri}. (27), logarithmic loss function, has the unique feature of penalizing volatility forecasts asymmetrically in low and high volatility periods as pointed out in \cite{Juri}.

To test whether the outcome is statistically significant or just an artifact of the dataset, we apply the Diebold-Mariano test to evaluate significance of improvement by introducing regime switching. For a particular loss function, we generate the time series of differences between the loss function values of portfolio volatility from two different covariance forecast models. Let the difference series be $d_{t}, t=1,..,N$. We would like to test the null hypothesis $H_{0}: E(d_t)=0$. On $\tau$-period-ahead forecasting horizon, the Diebold-Mariano statistics, taking into account of autocorrelation, is found to be:
\[DM=\frac{\bar{d}}{\sqrt{\frac    {   \hat{\gamma}_{d}(0) + 2\sum_{k=1}^{\tau-1}\hat{\gamma}_{d}(k)   }   {n}}} \quad \sim \quad N(0,1)\]
where 
\[  \hat{\gamma}_{d}(k)=\frac{1}{n} \sum_{t=|k|+1}^{n} (d_t-\bar{d})(d_{t-|k|}-\bar{d})  \]
and 
\[\bar{d}=\sum_{t=1}^{n}d_t\]

\section{Empirical results}
\subsection{Data}
\begin{figure}[H]
	\centering
	\begin{subfigure}{\textwidth}
		\centering
		\includegraphics[height=5.1cm,width=\linewidth]{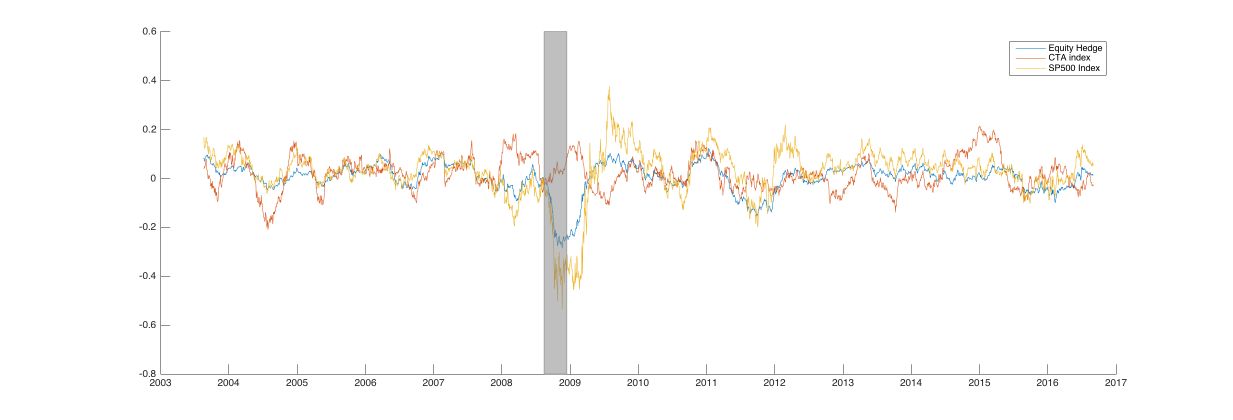}
	\end{subfigure}
	\caption{100-day simple moving average of asset returns}
	\label{fig:rowdataret}
	\vspace{0.5em}
	\caption*{*This plot shows 100-day simple moving average of returns of the three underlying assets. Shaded area corresponds to defined crisis period from 2008-08-15 to 2008-12-15.}
\end{figure}

The data set analyzed in this paper consists of Equity Hedge index (representing equity hedge fund manager), a SG Trend Index (representing trend followers) and an S\&P 500 index (representing the market). The sample period is from 2003-04-01 to 2016-09-01 for a total of 3381 observations. The first 900 observations (from 2003-04-01 to 2006-10-24) are used as in-sample period for estimation purposes while the remaining 2481 observations are taken as out-of-sample period for forecasting evaluation purposes. We define the crisis period as from 2008-08-15, one month before Lehman Brother filed bankruptcy, to 2008-12-15, two month after Lehman Brother filed bankruptcy. As can be seen from Figure \ref{fig:rowdataret}, this three-month crisis period is selected to capture S\&P500 downward trend in 2008 financial crisis. In our case, as we believe MRSOGARCH is more rapid in reflecting market condition changes, it is more insightful to test GMVP performance under a short and intense bear market rally as opposed to a prolonged period. Economic statistics of the three underlying assets over pre-crisis (2003-04-01 to 2008-08-15), crisis (2008-08-15 to 2008-12-15) and post-crisis period (2008-12-15 to 2016-09-01) are shown in table \ref{ehstats}-\ref{spxstats}.

\begin{table}[H]
	\centering
	\vspace{0.5em}
	\resizebox{0.8\textwidth}{!}{%
		\begin{tabular}{@{}llllll@{}}
			\toprule
			&Mean p.a.          & St. Dev. p.a. & 5\% Quantile & Worst Case & Sharpe Ratio             \\ \midrule
			pre-crisis period  & 4.364\%       & 5.860\%      & -0.635\%   & -1.833\%     & 0.74   \\
			crisis period      & -66.640\%     & 15.776\%     & -2.166\%   & -3.067\%     & -4.22 \\
			post-crisis period & 1.450\%       & 6.066\%      & -0.630\%   & -2.750\%     & 0.24   \\ \bottomrule
		\end{tabular}%
	}
	\vspace{0.5em}
	\caption{Risk Measures based on Equity Hedge Index}
	\label{ehstats}
	\vspace{0.5em}
	\resizebox{0.8\textwidth}{!}{%
		\begin{tabular}{@{}llllll@{}}
			\toprule
			&Mean p.a.          & St. Dev. p.a. & 5\% Quantile & Worst Case & Sharpe Ratio            \\ \midrule
			pre-crisis period  & 5.833\%       & 13.622\%     & -1.440\%   & -4.762\%    & 0.43  \\
			crisis period      & 38.368\%      & 13.145\%     & -1.202\%   & -3.025\%     & 2.92 \\
			post-crisis period & 2.150\%       & 11.061\%     & -1.148\%   & -3.743\%     & 0.19  \\ \bottomrule
		\end{tabular}%
	}
	\vspace{0.5em}
	\caption{Risk Measures based on SG Trend Index}
	\label{ctastats}
	\vspace{0.5em}
	\resizebox{0.8\textwidth}{!}{%
		\begin{tabular}{@{}llllll@{}}
			\toprule
			&Mean p.a.          & St. Dev. p.a. & 5\% Quantile & Worst Case & Sharpe Ratio             \\ \midrule
			pre-crisis period  & 7.916\%       & 13.807\%     & -1.409\%   & -3.534\%     & 0.57   \\
			crisis period      & -117.941\%    & 63.415\%     & -7.192\%   & -9.470\%     & -1.85 \\
			post-crisis period & 11.715\%      & 17.808\%     & -1.826\%   & -6.896\%     & 0.65   \\ \bottomrule
		\end{tabular}%
	}
	\vspace{0.5em}
	\caption{Risk Measures based on S\&P500 Index}
	\label{spxstats}
	\vspace{0.5em}
	\caption*{*Tables \ref{ehstats}-\ref{spxstats} show economic statistics of the three underlying assets. Mean p.a. and St.Dev.p.a. are mean annulized return and standard deviation in corresponding period. 5\% quantile shows quantiles for the cumulative probability 5\% of daily returns in corresponding period. Worst case is the worst daily return in corresponding period.}
\end{table}

From table \ref{ehstats}-\ref{ctastats}, we can see that in pre-crisis period Equity Hedge index has the most stable returns with pre-crisis standard deviation twice lower than standard deviation of both SG Trend Index and S\&P 500 index. Therefore, it can be expected that a GMVP will invest most heavily in this asset pre-crisis. In contrast, SG Trend Index is almost as volatile as S\&P 500 index prior to the crisis. However, during the financial crisis, we see significant change in asset return dynamics. Most notably, both equity hedge and S\&P500 index become highly volatile and suffered huge losses whereas SG Trend Index raises significantly in value and become less volatile. This reversal of asset return regime will require investors to adjust their portfolio upon arrival of the crisis so that more wealth can be invested into the SG Trend Index. For example, as equity hedge was more volatile than SG Trend Index during the crisis, we expect portfolio weights of SG Trend Index surpasses that of the equity hedge. If our claim that MRSOGARCH models are more flexible in accommodating sudden changes is true, we should observe that MRSOGARCH is indeed faster to allocate wealth in SG Trend Index. Moreover, as SG Trend Index maintains a positive mean return during the crisis, we expect that the model that is the fastest in recognizing the crisis also suffers the least amount of loss during the crash.

\subsection{Estimation}
Model estimation is based on three principal components from 2003-04-01 to 2006-10-24. Parameter estimation results are shown in tables \ref{parammrs}-\ref{paramgrch}. Here we see that the addition of regime switching is worthwhile from an in-sample point of view. By introducing regime switching, we document an increase fit of 16.3, 18.9, and 10.2 for the 1st, 2nd, 3rd principal component respectively. The significance of outperformance is tested with likelihood ratio test for each component. p-value from the test is 1.6318e-05, 1.2292e-06, and 0.0023 for the 1st, the 2nd, and the 3rd principal component respectively. Therefore the increase in goodness of fit is significant for all principal components.
\begin{table}[H]
	\centering
	\resizebox{0.41\textwidth}{!}{%
		\begin{tabular}{@{}lllll@{}}
			
			\toprule
			&$Y_{1}$        & $Y_{2}$                      & $Y_{3}$                                                    \\ \midrule
			$\omega^{(1)}$ & \makecell{0.0923\\(0.1388)}  & \makecell{1.3985\\(1.6880)}  & \makecell{0.0205\\(0.0043)}  \\\\ 
			$\omega^{(2)}$ & \makecell{0.0780\\(0.3127)}  & \makecell{0.0478\\(0.0314)}  & \makecell{0\\(0.0204)}       \\\\ 
			$\alpha^{(1)}$ & \makecell{0\\(0.0403)}  & \makecell{0\\(0.0556)}  & \makecell{0.0343\\(0.0220)}  \\\\ 
			$\alpha^{(2)}$ & \makecell{0.2058\\(0.1097)}  & \makecell{0\\(0.0889)}  & \makecell{0.9305\\(0.4160)}  \\\\ 
			$\beta^{(1)}$  & \makecell{0.7396\\(0.1092)}  & \makecell{0\\(1.2497)}  & \makecell{0.8329\\(0.0264)}  \\\\ 
			$\beta^{(2)}$  & \makecell{0.7417\\(0.1913)}  & \makecell{0.7290\\(0.0886)}  & \makecell{0\\(0.0851)}       \\\\ 
			p              & \makecell{0.8134\\(0.0819)}  & \makecell{0.9304\\(0.0363)}  & \makecell{0.9840\\(0.0054)}  \\\\ 
			q              & \makecell{0.5838\\(0.1694)}  & \makecell{0.9080\\(0.0328)}  & \makecell{0\\(0.0482)}  
			\\\\
			logLike        & -1496.3                       & -1227.1                       & -626.4                        \\ \bottomrule 
		\end{tabular}%
	}
	\vspace{0.5em}
	\caption{Parameter Estimation for MRSOGARCH}
	\label{parammrs}
	\vspace{0.5em}
\end{table}
\vspace{2em}
\begin{table}[H]
	\centering
	\resizebox{0.41\textwidth}{!}{%
		\begin{tabular}{@{}lllll@{}}
			\toprule
			&$Y_{1}$        & $Y_{2}$                     & $Y_{3}$                                                  \\ \midrule
			$\omega$ & \makecell{0.0856\\(0.045)}  & \makecell{0.017\\(0.0076)}  & \makecell{0.0078\\(0.003)}  \\\\
			$\alpha$ & \makecell{0.0556\\(0.0396)} & \makecell{0.0608\\(0.0172)} & \makecell{0.0614\\(0.022)}  \\\\
			$\beta$  & \makecell{0.8942\\(0.019)}  & \makecell{0.9233\\(0.014)}  & \makecell{0.9077\\(0.0147)} \\\\
			logLike        & -1512.6                     & -1246                       & -636.6                      \\ \bottomrule
		\end{tabular}%
	}
	\vspace{0.5em}
	\caption{Parameter Estimation for classic OGARCH}
	\label{paramgrch}
	\vspace{1em}
	\caption*{*Tables above show results of parameter estimation for principal components $Y_{1}$, $Y_{2}$, and $Y_{3}$. Standard errors of the parameter estimates are shown in brackets. logLike denotes optimized loglikelihood objective function value. Refer to section 2 for parameter notations.}
\end{table}

In his study of modeling exchange rates with regime switching GARCH, \cite{Klaassen} attributes better estimation fit of regime-switching GARCH to the allowance of temporary switch from high to low volatility regime. The two-regime construct allows regime switching GARCH to quickly adapt to and revert back from temporary changes in the return series such as large shocks and market crashes. Here we see empirical evidence that GARCH’s inflexibility can be improved by introducing one additional regime. This comes naturally as principal components characterize sudden moves in original return series, e.g. large shocks and sudden crashes. Better estimation fit, though not direct evidence of better forecasting performance, does show that the principal components are better described by a regime-switching GARCH model. In other words, better in-sample estimation evidenced our claim that regime-switching would enhance models’ adaptability to market changes by introducing one extra source of flexibility. As OGARCH model relies on univariate GARCH’s modeling of principal components, we expect and will later show that incorporation of regime-switching will improve OGARCH’s performance overall.

\subsection{Forecasting and portfolio selection}
So far we have demonstrated that introducing regime-change GARCH indeed improved classic OGARCH from an estimation point of view. However, as the improvement is shown merely on estimation of principal components, it does not serve as direct evidence of better forecasting. In this section, we investigate the quality of covariance forecasts in a practical context. As discussed in previous section, a GMVP is constructed based on each model's covariance matrix forecasts. Portfolio performance out-of-sample is investigated on both one-day and five-day investment horizons. In addition to basic statistics, we will also use Sharpe ratio to gauge the risk-reward trade-off.
\[\text{Sharpe Ratio}:\quad SR=\frac{\text{annualized rate of return}}{\text{annualized volatility}}\]
But note that as GMVPs only look to minimize portfolio variance, evaluation of model forecasting performance should be based on portfolio volatility instead of return. 

One key argument we put forth in previous sections is that, as return volatility of equity hedge and S\&P 500 index soar upon 2008 market crash, MRSOGARCH's more flexible structure should allow it to recognize change of market regime faster and thus to instruct its GMVP to allocate wealth in SG Trend Index more promptly. To investigate this, (1) we compare standard deviation of portfolio returns during crisis as shown in Table \ref{econstats1}-\ref{econstats4}; (2) we plot GARCH(1,1) inferred daily portfolio variance around the crisis period to see if MRSOGARCH indeed lowers portfolio variance upon the market crash; (3) we plot portfolio weights around crisis period to see if MRSOGARCH is indeed faster in allocating wealth to SG Trend Index; (4) we plot mean portfolio return in Figure \ref{ret1001d}-\ref{ret1005d} to see if MRSOGARCH suffers less loss by faster recognizing the crisis; (5) we plot smoothed probability of high-variance state from first principle component estimates to see if MRSOGARCH indeed recognized market crash by assigning higher probability to high-variance regime.

We also assess models' overall forecasting power by measuring their predictive accuracy. As discussed in section 3, predictive accuracy of the forecasting models is measured using five loss functions. The idea is to construct an equal-weight portfolio and compare the ``observed portfolio volatility'' the portfolio volatility forecasts, where the absolute value of realized portfolio returns over each investment period are taken as “observed volatility” and $h_{p,t+1}=\sqrt{\omega_{eq}^{T} \Sigma_{t+\tau} \omega_{eq}}$ are taken as portfolio volatility forecasts. In this study, we assess model predictive accuracy based on both 1-day and 5-day holding portfolios. The statistical significance of outperformance is tested with Diebold-Mariano test.

\begin{table}[H]
	\centering
	\resizebox{0.9\textwidth}{!}{%
		\begin{tabular}{@{}llllllll@{}}
			\toprule
			&Mean Ret. P.a. & Std. Dev. P.a & 5\% Quantile & Worst case & Max Draw Down & Sharpe Ratio &           \\ \midrule
			MRSOGARCH      & 0.3402\%      & 6.0186\%     & -0.6638\%  & -2.0431\%     & -19.4174\%   & 0.5455\%  \\
			OGARCH         & 0.0829\%      & 6.0840\%     & -0.6641\%  & -2.0102\%     & -21.6199\%   & 0.2777\%  \\
			EWMA           & -0.5834\%     & 6.3215\%     & -0.7110\%  & -2.1598\%     & -25.1040\%   & -0.3837\% \\ \bottomrule
		\end{tabular}%
	}
	\vspace{0.3em}
	\caption{Portfolio statistics 1-day investment horizon on entire period}
	\label{econstats1}
	\vspace{0.5em}
	
	\resizebox{0.9\textwidth}{!}{%
		\begin{tabular}{@{}llllllll@{}}
			\toprule
			&Mean Ret. P.a. & Std. Dev. P.a & 5\% Quantile & Worst case & Max Draw Down & Sharpe Ratio &           \\ \midrule
			MRSOGARCH      & 0.0073\%      & 6.0798\%     & -0.6717\%  & -2.1869\%     & -21.4752\%   & 0.1994\%  \\
			OGARCH         & -0.1293\%     & 6.2048\%     & -0.6865\%  & -2.1000\%     & -22.6727\%   & 0.0644\%  \\
			EWMA           & -0.7172\%     & 6.3846\%     & -0.7222\%  & -2.2598\%     & -25.6337\%   & -0.5088\% \\ \bottomrule
		\end{tabular}%
	}
	\vspace{0.3em}
	\caption{Portfolio statistics 1-week investment horizon on entire period}
	\label{econstats2}
	\vspace{0.5em}
	
	\resizebox{0.9\textwidth}{!}{%
		\begin{tabular}{@{}llllllll@{}}
			\toprule
			&Mean Ret. P.a. & Std. Dev. P.a & 5\% Quantile & Worst case & Max Draw Down & Sharpe Ratio &            \\ \midrule
			MRSOGARCH      & -19.0723\%    & 7.2101\%     & -1.1429\%  & -1.5765\%     & -8.2001\%    & -18.3640\% \\
			OGARCH         & -27.3365\%    & 8.4884\%     & -1.1930\%  & -2.0102\%     & -12.0406\%   & -23.5568\% \\
			EWMA           & -35.1514\%    & 10.2580\%    & -1.3718\%  & -2.1209\%     & -14.8917\%   & -26.4101\% \\ \bottomrule
		\end{tabular}%
	}
	\vspace{0.3em}
	\caption{Portfolio statistics 1-day investment horizon on crisis period}
	\label{econstats3}
	\vspace{0.5em}
	
	\resizebox{0.9\textwidth}{!}{%
		\begin{tabular}{@{}llllllll@{}}
			\toprule
			&Mean Ret. P.a. & Std. Dev. P.a & 5\% Quantile & Worst case & Max Draw Down & Sharpe Ratio &            \\ \midrule
			MRSOGARCH      & -27.1703\%    & 8.2262\%     & -1.1462\%  & -1.7501\%     & -11.8632\%   & -24.1489\% \\
			OGARCH         & -31.9653\%    & 9.6942\%     & -1.3523\%  & -2.1000\%     & -13.6650\%   & -24.8524\% \\
			EWMA           & -37.6295\%    & 10.7456\%    & -1.4333\%  & -2.2598\%     & -15.5907\%   & -27.4754\% \\ \bottomrule
		\end{tabular}%
	}
	\vspace{0.3em}
	\caption{Portfolio statistics 1-week investment horizon on crisis period}
	\label{econstats4}
	\vspace{0.5em}
	\caption*{*Tables above show GMVP statistics for the three models on 1-day and 5-day investment horizon. Mean Ret. P.a. denotes annualized portfolio mean return over corresponding period. Std. Dev. P.a. denotes annualized standard deviation of portfolio return over corresponding period. 5\% quantile shows quantiles for the cumulative probability 5\% of daily returns in corresponding period. Worst case is the worst daily return in corresponding period. Max Draw Down is the maximum of the decline from a historical peaks over the corresponding period. }
\end{table}

\begin{figure}[H]
	\centering
	\begin{subfigure}{\textwidth}
		\centering
		\includegraphics[height=5.1cm,width=\linewidth]{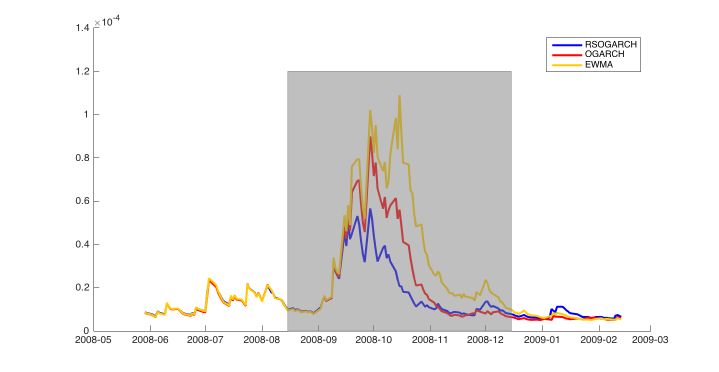}
	\end{subfigure}
	\caption{Inferred portfolio variance on 1-day investment horizon from garch(1,1)}
	\label{garchvard1}
	
	\begin{subfigure}{\textwidth}
		\centering
		\includegraphics[height=5.1cm,width=\linewidth]{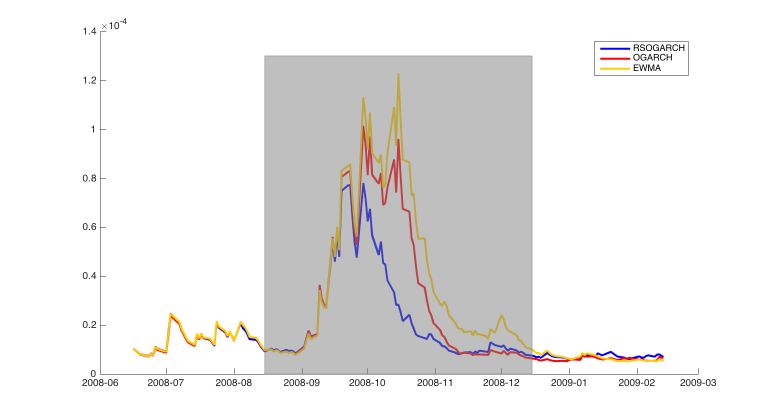}
	\end{subfigure}
	\caption{Inferred portfolio variance on 5-day investment horizon from garch(1,1)}
	\label{garchvard5}
	\vspace{0.5em}
	\caption*{*Figures \ref{garchvard1}-\ref{garchvard5} show GARCH(1,1) inferred portfolio variance around crisis period. Shaded area highlights the crisis period as defined in section 4.1.}
\end{figure}

\begin{figure}[H]
	\centering
	\begin{subfigure}{\textwidth}
		\centering
		\includegraphics[height=5.1cm,width=0.9\linewidth]{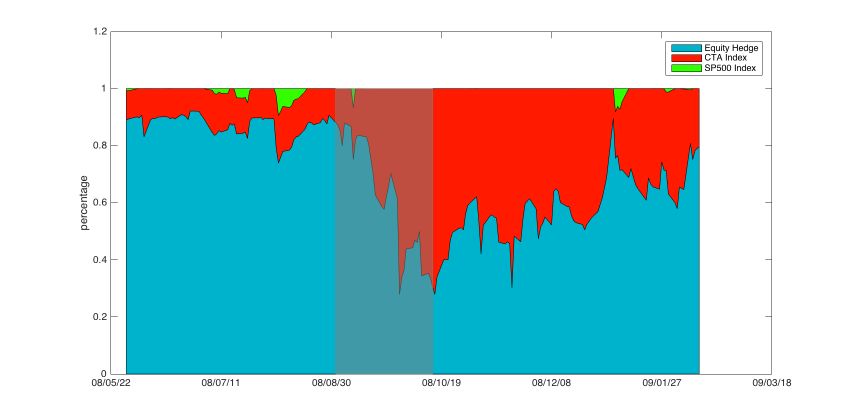}
		\label{fig:sub1}
	\end{subfigure}
	\caption{MRSOGARCH GMVP weights around crisis period}
	\label{weightmrs1}
	
	\centering
	\begin{subfigure}{\textwidth}
		\centering
		\includegraphics[height=6cm,width=0.9\linewidth]{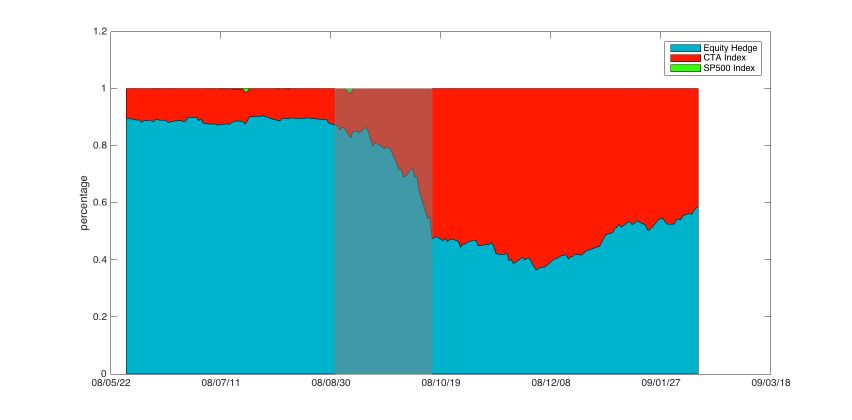}
		\label{fig:sub1}
	\end{subfigure}
	\caption{OGARCH GMVP weights around crisis period}
	\label{weightorg1}
	
	\centering
	\begin{subfigure}{\textwidth}
		\centering
		\includegraphics[height=6cm,width=0.9\linewidth]{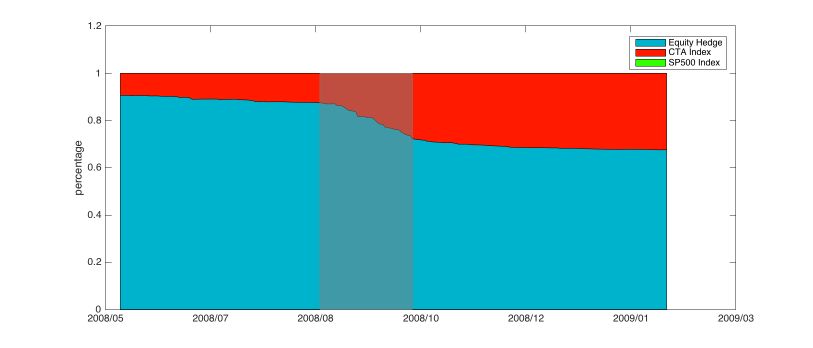}
		\label{fig:sub1}
	\end{subfigure}
	\caption{EWMA GMVP weights around crisis period}
	\label{weightew1}
	\vspace{0.5em}
	\caption*{*Figures \ref{weightmrs1}-\ref{weightew1} show GMVP weights around the crisis period. The shaded area highlights period from 2008-09-01 to 2008-10-15, i.e. from one week before Lehman Brother bankruptcy to one month after Lehman Brother bankruptcy.}
\end{figure}

\begin{figure}[H]
	
	\centering
	\begin{subfigure}{\textwidth}
		\centering
		\includegraphics[height=6cm,width=0.9\linewidth]{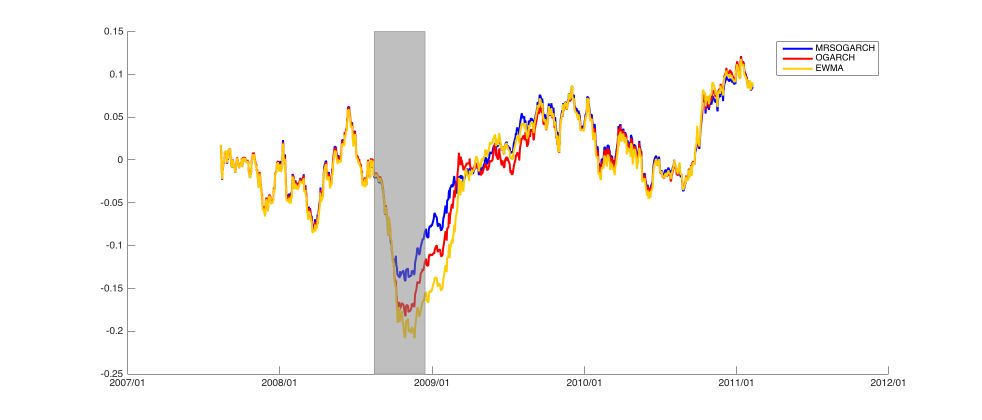}
		\label{fig:sub1}
	\end{subfigure}
	\caption{100-day rolling average of portfolio returns on 1-day investment horizon}
	\label{ret1001d}
	
	\centering
	\begin{subfigure}{\textwidth}
		\centering
		\includegraphics[height=6cm,width=0.9\linewidth]{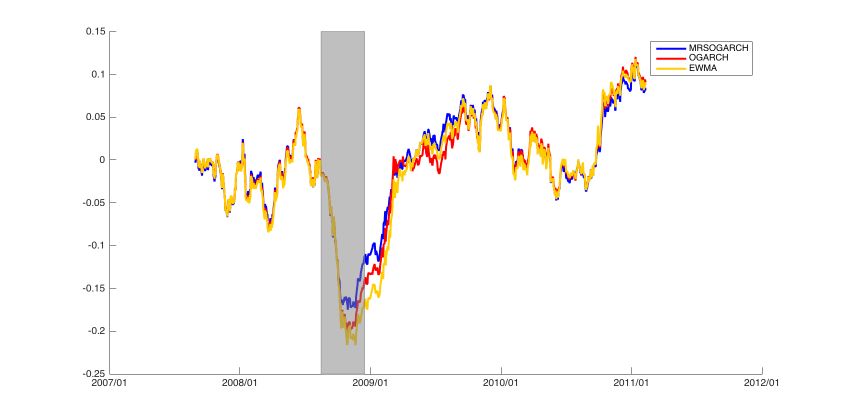}
		\label{fig:sub1}
	\end{subfigure}
	\caption{100-day rolling average of portfolio returns on 5-day investment horizon}
	\label{ret1005d}
	\vspace{0.5em}
	\caption*{*Figures \ref{ret1001d}-\ref{ret1005d} show 100-day rolling average of portfolio returns on 1-day and 5-day investment horizon. The shaded area highlights crisis period defined in section 4.1.}
\end{figure}

\begin{figure}[H]
	\centering
	\begin{subfigure}{\textwidth}
		\centering
		\includegraphics[height=12cm,width=0.9\linewidth]{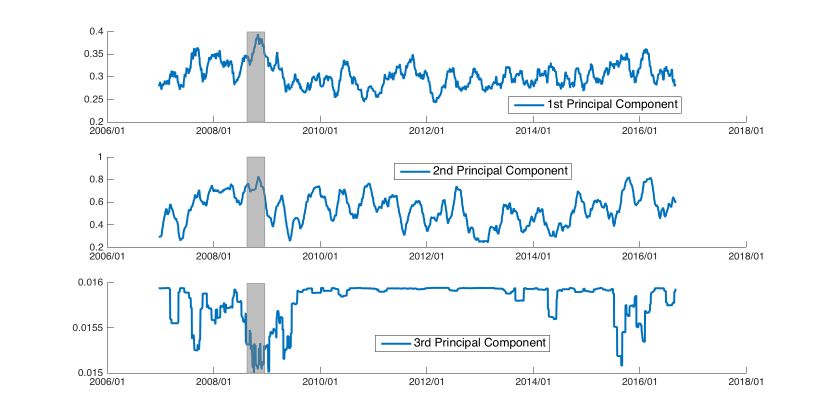}
		\label{fig:sub1}
	\end{subfigure}
	\caption{50-day rolling average of probability in state 1}
	\label{ppmarked}
	\vspace{0.5em}
	\caption*{*Figure \ref{ppmarked} shows 50-day simple rolling average of inferred probability (smoothed) in state 1 from the first, second and third principal component estimation as of the most recent day. Shaded area highlights crisis period as defined in section 2.1.}
\end{figure}

\begin{table}[H]
	\centering
	\resizebox{\textwidth}{!}{%
		\begin{tabular}{@{}llllllll@{}}
			\toprule
			& \multicolumn{3}{l}{Holding period: 1-day} &  & \multicolumn{3}{l}{Holding period: 1-week}      \\ \midrule
			&MRSOGARCH     & OGARCH       & EWMA        &  & MRSOGARCH              & OGARCH    & EWMA                 \\
			$MSE_1$  & 1.61e-05 & 1.82e-05 & 1.85e-05 & & 7.58e-05 & 8.80e-05 & 7.81e-05 \\ 
			$MSE_1$  & 3.64e-05 & 4.03e-05 & 3.71e-05 & & 1.66e-04 & 1.91e-04 & 1.65e-04 \\
			$MAD_1$  & 3.02e-03 & 3.38e-03 & 3.10e-03 & & 6.45e-03 & 7.35e-03 & 6.50e-03 \\
			$MAD_2$  & 3.64e-05 & 4.03e-05 & 3.71e-05 & & 1.66e-04 & 1.91e-04 & 1.65e-04 \\
			$R2LOG$ & 6.40     & 7.35     & 6.47     & & 5.00     & 6.08     & 5.15     \\ \bottomrule
		\end{tabular}%
	}
	\vspace{0.3em}
	\caption{loss functions for one-day, one-week ahead forecast horizon}
	\label{lossnum}
	\vspace{0.5em}
	\caption*{*Table \ref{lossnum} shows deviation of volatility forecast of an equal-weight portfolio from its daily returns (used as a proxy to ``real'' volatility) under three competing models evaluated with loss functions listed in section 3.}
	\vspace{1.2em}
	
	\centering
	\resizebox{0.85\textwidth}{!}{%
		\begin{tabular}{@{}llll@{}}
			\toprule
			&MRSOGARCH vs OGARCH & MRSOGARCH vs EWMA       & OGARCH vs EWMA                                                 \\ \midrule
			$MSE_1$ & \makecell{9.4559\\(0)}  & \makecell{4.3896\\(5.6779e-06)} & \makecell{0.41888\\(0.33765)}    \\\\ 
			$MSE_1$ & \makecell{13.0063\\(0)} & \makecell{1.8085\\(0.035262)}   & \makecell{-5.8505\\(2.4506e-09)} \\\\
			$MAD_1$ & \makecell{17.625\\(0)}  & \makecell{2.9066\\(0.001827)}   & \makecell{-8.1452\\(2.2204e-16)} \\\\
			$MAD_2$ & \makecell{13.0063\\(0)} & \makecell{1.8085\\(0.035262)}   & \makecell{-5.8505\\(2.4506e-09)} \\\\
			$R2LOG$ & \makecell{23.1306\\(0)} & \makecell{1.7267\\(0.042115)}   & \makecell{-15.3376\\(0)}         \\ \bottomrule
		\end{tabular}%
	}
	\vspace{0.3em}
	\caption{Pairwise comparison among competing models by Diebold Mariano test at 1-day investment horizon}
	\label{dm1d}
	\vspace{1em}
	
	\centering
	\resizebox{0.85\textwidth}{!}{%
		\begin{tabular}{@{}llll@{}}
			\toprule
			&MRSOGARCH vs OGARCH & MRSOGARCH vs EWMA               & OGARCH vs EWMA                                                  \\ \midrule
			$MSE_1$ & \makecell{4.8279\\(6.899e-07)}  & \makecell{0.72063\\(0.23557)}  & \makecell{-2.0655\\(0.019436)}   \\\\
			$MSE_1$ & \makecell{6.3763\\(9.0736e-11)} & \makecell{-0.24733\\(0.40233)} & \makecell{-4.3218\\(7.738e-06)}  \\\\
			$MAD_1$ & \makecell{7.0219\\(1.0941e-12)} & \makecell{0.41964\\(0.33737)}  & \makecell{-5.0752\\(1.9355e-07)} \\\\
			$MAD_2$ & \makecell{6.3763\\(9.0736e-11)} & \makecell{-0.24733\\(0.40233)} & \makecell{-4.3218\\(7.738e-06)}  \\\\
			$R2LOG$ & \makecell{10.2186\\(0)}         & \makecell{1.8581\\(0.031576)}  & \makecell{-8.1547\\(2.2204e-16)} \\ \bottomrule
		\end{tabular}%
	}
	\vspace{0.3em}
	\caption{Pairwise comparison among competing models by Diebold Mariano test at 5-day investment horizon}
	\label{dm5d}
	\vspace{0.5em}
	\caption*{*Table \ref{dm1d}-\ref{dm5d} show test statistics and p-values from Diebold Mario tests at 1-day and 5-day investment horizon. p-values are enclosed within the brackets.}
\end{table}

Tables \ref{econstats1}-\ref{econstats4} report GMVP performance under EWMA, MRSOGARCH and OGARCH correlation/volatility forecasts. As GMVP only looks to minimize investment risk, model with better correlation/volatility forecasts will mostly likely generate most stable returns. We notice that MRSOGARCH GMVP's returns are indeed least volatile over entire and crisis period under both 1-day and 5 day investment horizon. Previously we argued that MRSOGARCH is able to generate better forecasts because it is more flexible in accommodating sudden market condition changes with an extra regime. This claim can be verified by comparing magnitude of outperformance that MRSOGARCH has over OGARCH and EWMA over entire and crisis period in terms of standard deviation of portfolio returns. Under 1-day investment horizon, MRSOGARCH is able to outperform OGARCH and EWMA by 0.07\% and 0.24\% over the entire period whereas the outperformance over OGARCH and EWMA is enlarged to 1.3\% and 3\% over crisis period. The same pattern can be observed for 5-day investment horizon. Plots of GARCH inferred GMVP portfolio variance also support this claim. In Figures \ref{garchvard1}-\ref{garchvard5}, we observe that as the GMVP volatility shoots up due to 2008 market crash, MRSOGARCH GMVP volatility is significantly lower on both 1-day and 5-day ahead forecasting horizon. Notice that GMVP volatility is much closer for the three competing models under tranquil market condition. This supports our claim that MRSOGARCH is much faster in recognizing the dramatic market breaks.

A particular point of interest then is whether MRSOGARCH indeed shows faster recognition of the crisis and makes corresponding adjustment to its portfolio more promptly. Figures \ref{weightmrs1}-\ref{weightew1} plot portfolio asset allocation based on MRSOGARCH, OGARCH and EWMA forecasts over one-day investment horizon (weight allocation over one-week investment horizon follow similar pattern). Apparently, both OGARCH and MRSOGARCH focus mainly on Equity Hedge and SG Trend Index before the market crash. One week prior to Lehman Brother Bankruptcy (2008-09-01), we observe that the two hold very similar portfolio: MRSOGARCH invests 90.7\% into Equity Hedge and 9.3\% into SG Trend Index while classic OGARCH invests 87.9\% into Equity Hedge and 12.1\% into SG Trend Index. From this point on until two weeks after Lehman Brother bankruptcy (2008-10-15), we observe that both portfolios undergo significant restructuring. Both portfolios reduced their holding in Equity Hedge. Notably, MRSOGARCH indeed demonstrates faster crisis recognition in this process and respond more promptly to the crisis by increasing its investment in SG Trend Index. From 2008-09-01 till 2008-10-15, average exposure of MRSOGARCH to Equity Hedge and SG Trend Index is 63.2\% and 36.6\% whereas OGARCH has a much higher exposure to Equity Hedge at 77.1\% with only 28.8\% invested in SG Trend Index). MRSOGARCH on average reduces 1.9\% of its investment in Equity Hedge per day whereas the rate for OGARCH is only 1.1\%. It is also visually obvious by comparing Figure \ref{weightmrs1} and Figure \ref{weightorg1} that MRSOGARCH is able to react to crisis much faster by allocating its wealth into SG Trend Index. Similarly, we observe faster reaction of MRSOGARCH to the market downturn on 1-week investment horizon. From 2008-09-01 to 2008-10-15, average exposure of MRSOGARCH to Equity Hedge and SG Trend Index is 58.7\% and 41.3\% while classic OGARCH has 71.2\% exposure in Equity Hedge and only 28.8\% in SG Trend Index. MRSOGARCH on average reduce 1.7\% of its investment in Equity Hedge per day whereas the rate for OGARCH is only 1.3\%. 

By portfolio rebalancing over the crisis period, MRSOGARCH is able to push Equity Hedge out of portfolio rapidly and therefore are able to avoid excessive losses from Equity Hedge’s poor performance over the crisis. In comparison, classic OGARCH reacts to the market crash more reluctantly and therefore suffers a larger downturn due to its exposure to Equity Hedge. From Figures \ref{ret1001d}-\ref{ret1005d} we observe that MRSOGARCH GMVP is able to reverse loss faster than OGARCH GMVP. The trend we observe Figures \ref{ret1001d}-\ref{ret1005d} is also reflected in other risk measures including maximum drawdown, worst-case return, and 5\% quantile returns. As shown in Table \ref{econstats3}-\ref{econstats4}, all these risk measures favor MRSOGARCH as the better alternative to avoid large investment loss than OGARCH. Notably, outperformance of MRSOGARCH GMVP is less pronounced over 5-day investment horizon due to ``reversion to the mean'' effect. Namely, with longer forecasting horizon and less frequent portfolio rebalancing, it is harder for MRSOGARCH GMVP to outperform as significantly. 

Previously we argued that the better performance of regime-switching OGARCH is due to the extra source of volatility persistence it introduced to model principal components’ time series. Figure \ref{ppmarked} shows the plot of 50-day simple moving average of smoothed high-variance-regime probability (as of the most recent day). The plot reflects the model's belief of the current market condition. We observe that all three principal components have undergone ``transition'' of regime to a certain degree. Notice that during the crisis period, the inferred state 1 probability of the first and second principal components surge to all-time high while that of the third component's plummets considerably. The fact that all three principal components' state 1 smoothed probability undergoes large changes during the crisis indicates that the model has indeed recognized the crisis and adjusted its estimation/forecasting accordingly. By adjusting regime probability, the model essentially lends more weight to the high-variance regime and therefore is able to reduce the delaying effect and to react to the crisis more quickly. 

The results of predictive accuracy evaluation based on the five loss functions are shown in table \ref{lossnum}. Pairwise comparison between competing models by Diebold Mariano test is shown in tables \ref{dm1d}-\ref{dm5d} for one-day portfolio holding and one-week portfolio holding. This is basically to compare how well the three competing models' forecasts match the actual outcome through an artificially constructed equal-weight portfolio. MRSOGARCH outperforms both OGARCH and EWMA significantly based on all five loss functions. This is quite natural considering that MRSOGARCH is more robust in adapting to abrupt market changes. Given the high significance of outperformance (small p-value) across the five loss functions, we have shown that MRSOGARCH is indeed superior to OGARCH and EWMA in terms of predictive accuracy. Results of the predictive accuracy evaluation largely echoes findings from the GMVP economic performance. As an example, the outperformance of MRSOGARCH over the other two models over 5-day investment horizon is less prominent compared to the results of one-day portfolio holding. This indicates that MRSOGARCH's advantage in terms of predictive accuracy decreases as the forecasting horizon lengthens. This finding aligns with \cite{Juri}'s experiment in the univariate setting. Meanwhile, we observe that GMVP performance for MRSOGARCH is less impressive on 5-day forecasting horizon. The alignment between model predictive accuracy and models’ economic performance reassures our claim that the model generates most accurate forecasts should form portfolio that best aligns with the investment goal.

\section{Conclusion}
Based on our methodology, we are able to construct a regime-switching model capable of modeling both correlation and volatility of multivariate time series. It has been shown via toy examples that for short forecasting horizon the proposed MRSOGARCH model is able to adjust more promptly to changing ``covariance structures'' while persevering the nice dimension reduction property of OGARCH. In the Fund of Funds management example, We have demonstrated that by reallocating wealth more promptly into SG Trend Index during stressed period, MRSOGARCH outperforms classic OGARCH model and EWMA model substantially in realized portfolio return and return volatility. Predictive accuracy of the proposed model is evaluated with several loss functions. The loss function values are compared to OGARCH and EWMA and it is concluded from pair-wise DM test that the outperformance is highly significant. This method retains advantage of classic OGARCH to allow applications to large number of sub-assets while it enhances OGARCH’s flexibility to accommodate sudden market changes. 

\bibliography{Manuscript}
\bibliographystyle{plainnat}

\end{document}